\documentstyle[emulateapj,pstricks]{article}

\newcommand{\DD}{{\mathcal{D}}}
\newcommand{\EE}{{\mathcal{E}}}

\begin{document}

\title{The Trispectrum of the 4 Year $COBE$-DMR data}

\author {M.~Kunz\altaffilmark{1}, A.J.~Banday\altaffilmark{2}, 
P.G.~Castro\altaffilmark{1}, 
  P.G.~Ferreira\altaffilmark{1}, K.M.~G\'{o}rski\altaffilmark{3,4}}
\altaffiltext{1}{Astrophysics, University of Oxford, Keble Road, 
Oxford OX1 3RH, UK}
\altaffiltext{2}{Max Planck Institut fuer Astrophysik, Karl-Schwarzschildstr.
 1,  Postfach 13 17, D-85741 Garching, Germany}
\altaffiltext{3}{European Southern Observatory, Garching, Germany}
\altaffiltext{4}{Warsaw University Observatory, 
 Aleje Ujazdowskie 4, 00-478 Warszawa, Poland.}

\begin{abstract}
We propose an estimator for the trispectrum of a
scalar random field on a sphere, discuss its geometrical
and statistical properties, and outline its implementation. 
By estimating the trispectrum of the 4 year $COBE$-DMR data 
(in HEALPix pixelization) we find new evidence of a non-Gaussian
signal associated with a known systematic effect. 
We find that by removing data from the sky maps for those periods of 
time perturbed by this effect, the amplitudes of the trispectrum coefficients
become completely consistent with predictions for a Gaussian sky.
These results reinforce the importance of 
statistical methods based in harmonic space for quantifying
non-Gaussianity.
\end{abstract}


\section{Introduction}
The Cosmic Microwave Background (CMB) is the cleanest window
on the origin of structure in the very early universe. A complete
description of the statistical properties of cosmological fluctuations
at a redshift $z\simeq1000$ affords us an essential insight
into those processes which may have seeded the formation of
galaxies. 
In a Gaussian theory of structure formation, such as the currently favored
model of Inflation, the power spectrum contains all the possible information
about the fluctuations. Any higher
order moment can subsequently be described in terms of it. However, if
the theory is non-Gaussian (as expected for structure formation theories
due to local effects from primordial
phase transitions or more generally from non-linear processes),
then there will be deviations from the simple Gaussian expressions
for the higher order moments.
Such behavior can serve as a powerful discriminator
between different models of structure formation.

Most analyses of CMB data to-date have focused on the
angular power spectrum and its sensitivity to various parameters 
of cosmological theories.
Some work has been done on the estimation of the three
point correlation function and its analogue in spherical harmonic
space, with intriguing results (Heavens 1998, Ferreira, 
Magueijo and G\'{o}rski 1998, Magueijo 2000, Banday, Zaroubi 
and G\'{o}rski 2000). It is the purpose of this
letter to propose a method for estimating the four point 
spectrum, the {\it trispectrum}, and to apply it to the
$COBE$ 4 year DMR data. This work complements the recent
work of Hu (2001) where some of the properties of the angular
trispectrum of the CMB are discussed.

The outline of this letter is as follows. In Section \ref{est} we
construct a set of orthonormal estimators and describe their properties
for a Gaussian random field. In Section \ref{res} we apply the
estimators to the $COBE$ 4 year DMR data. We show that we
detect the non-Gaussian signal found in Ferreira, Magueijo $\&$ G\'{o}rski
(1998) and that it can be explained by the arguments presented
in Banday, Zaroubi \& G\'{o}rski (2000), and in particular that this
is a manifestation of a known systematic effect. 
We therefore conclude that
the $COBE$ 4 year data is consistent with a Gaussian cosmological signal.
In  Section \ref{dis} we summarize our results.

\section{The estimator}
\label{est}
In this section we wish to construct a set of quantities for
estimating the trispectrum of a random field on the sphere.
The temperature anisotropy in a given direction on the celestial
sphere, $T({\bf n})$, can be expanded in
terms of spherical harmonic functions, $Y_{\ell m}({\bf n})$:
\begin{equation}
  T({\bf n})=\sum_{\ell m}a_{\ell m}Y_{\ell m}({\bf n})
\end{equation}
For any theory of structure formation, the $a_{\ell m}$ coefficients are
a set of random variables; we shall restrict ourselves to
theories which are statistically homogeneous and isotropic.
In this case we can define the power spectrum $C_\ell$ of the
temperature anisotropies by $\langle a_{\ell m}
a^*_{\ell' m'}\rangle=C_{\ell}\delta_{\ell \ell'}$. 

We now seek to construct a set of tensors that are geometrically
independent, describe their statistical properties
for a Gaussian random field and then discuss the practical
issue of their implementation.
Given a set of $a_{\ell m}$ we wish to find the index structure of
the set of four point correlators such that (1) they are 
rotationally invariant (2) they form a complete basis 
(preferably orthonormal) of the whole space of admissible
four-point correlators and
(3) they satisfy the appropriate symmetries under interchanges of
$m$- and $\ell$-values. We shall restrict ourselves to the case in which
$\ell_1=\ell_2=\ell_3=\ell_4=\ell$. Furthermore, throughout this
section we keep $\ell$ fixed. We determine the tensor ${\cal T}$
such that
\begin{equation}
\langle a_{\ell m_1} a_{\ell m_2} a_{\ell m_3} a_{\ell m_4}\rangle
=\sum_{a=0}^{n}T_{\ell;a}{\cal T}^{a;\ell}_{m_1m_2m_3m_4} \label{4mom}
\end{equation}
where $n$=${\rm int}(\ell/3)$ (due to reflection, permutation and rotational
symmetry). The $T_{\ell;a}$ values are then the components of
the trispectrum which we wish to estimate.
The explicit form of the
${\cal T}$ are
\begin{eqnarray}
&&{\cal T}^{a;\ell}_{m_1m_2m_3m_4}=\sum_{\alpha=0}^{\ell}
{\cal L}^{a\alpha}_{\ell}
{\bar {\cal T}}^{\alpha;\ell}_{m_1m_2m_3m_4} \label{llmatrix} \\
&&{\bar {\cal T}}^{\alpha;\ell}_{m_1m_2m_3m_4}= 
\sum_{M=-2\alpha}^{2\alpha}(-1)^M
\left( \begin{array}{ccc} \ell & \ell & 2\alpha \\
        m_1 & m_2 & M \end{array} \right)\times\nonumber\\  
&&\left( \begin{array}{ccc} 2\alpha & \ell & \ell \\
        -M & m_3 & m_4 \end{array} \right) + \mbox{inequiv. permutations}
\label{tens}
\end{eqnarray}
where the matrices in parentheses are the Wigner 3-J symbols. The 
${\bar {\cal T}}^{\alpha;\ell}$ are not orthogonal and satisfy
\begin{eqnarray}
{\bar {\cal T}}^{\alpha;\ell}_{m_1m_2m_3m_4}{\bar {\cal T}}^{\alpha;\ell}_{m_1m_2m_3m_4}=\frac{3}{4\alpha+1}\delta_{\alpha\beta}+
6
\left\{ \begin{array}{ccc} \ell & \ell & 2\alpha \\
        \ell & \ell & 2\beta \end{array}  \right\} \nonumber
\end{eqnarray}
(where summation over the $m_i$ is assumed) which has rank $n+1$.
The matrix ${\cal L}_\ell$ in (\ref{llmatrix}) 
is a rectangular matrix (with a triangular
sub-block) with $n+1$ columns and $\ell+1$ rows. It is constructed 
through a Gram-Schmidt procedure by subtracting for each $\alpha$
(starting from $\alpha=0$)
the projection onto all $a'<a$ and then normalizing the result.
The $\alpha=0$ (and hence $a=0$) tensor is proportional to the
Gaussian contribution. 
This can be easily seen given that for $\alpha=0$ the Wigner 3J
symbols are simply Kronecker $\delta$ symbols in the corresponding indices.
The remaining $a>0$ terms contain therefore no Gaussian signal
and quantify the non-Gaussian part of the trispectrum.

The ${\cal T}$ are orthonormal and can be used to construct an 
estimator for $T_a$ from a realization of $a_{\ell m}$:
\begin{equation}
{\hat T}_{\ell;a}={\cal T}^{a;\ell}_{m_1m_2m_3m_4}a_{\ell m_1} a_{\ell m_2} a_{\ell m_3} a_{\ell m_4} \label{tnull}
\end{equation}
For a Gaussian random field we expect 
$\sigma^2[{\hat T}_{\ell;0}]\gg\sigma^2[{\hat T}_{\ell;a}]$ for
$a>0$, where $\sigma^2[A]$ denotes the variance of the random
variable $A$ and ${\hat T}_{\ell;0}$ is simply the square of the minimum
variance estimator of the $C_\ell$. One finds that 
$\langle{\hat T}_{\ell;a}\rangle=0$ and $\sigma^2[{\hat T}_{\ell;a}]=
24 C_\ell^4$ for all $a>0$. 

To show that the 
${\hat T}_{\ell;a}$ constitute a family of minimum variance estimators
we construct a linear combination of the estimators:
\begin{equation}
{\cal T}^{\ell}_{m_1m_2m_3m_4}=\sum_{a=0}^{n}c_a{\cal T}^{a;\ell}_{m_1m_2m_3m_4}
\end{equation}
and minimize the function
\begin{eqnarray}
\sigma^2_\ell[c_a,\lambda]=\langle({\cal T}^{\ell}_{m_1m_2m_3m_4}a_{\ell m_1} a_{\ell m_2} a_{\ell m_3} a_{\ell m_4})^2\rangle \nonumber \\
-\langle{\cal T}^{\ell}_{m_1m_2m_3m_4}a_{\ell m_1} 
a_{\ell m_2} a_{\ell m_3} a_{\ell m_4}\rangle^2 \nonumber \\
-\lambda C^2_\ell ({\cal T}^{\ell}_{m_1m_2m_3m_4}
{\cal T}^{\ell}_{m_1m_2m_3m_4}-1)
\end{eqnarray}
where summation over all $m_i$ is implied. The last term, a Lagrange 
multiplier, ensures that ${\cal T}^{\ell}$ is normalized. We solve
$\partial_{c_a} \sigma^2_\ell[c_a,\lambda]=\partial_\lambda \sigma^2_\ell[c_a,\lambda]=0$ 
to find a set of two equations
\begin{equation}
(24I+72A_\ell)^{ab}c_b+\lambda c_b=0 \quad \mathrm{and} \quad c^2=1
\end{equation}
where
$$
A^{ab}_\ell={\cal T}^{a;\ell}_{m_1m_2m_am_a}{\cal T}^{b;\ell}_{m_1m_2m_bm_b} .
$$
This is an eigenvector equation where, for a given eigenvector ${\bf c}$,
the eigenvalue $\lambda$ will give the expected variance of the
estimator. 
Of the $n+1$ eigenvalues, one is large and
has an eigenvector proportional to ${\hat T}_{\ell;0}$.
The remaining eigenvalues
have an amplitude $\lambda=24$ and each eigenvector is
a ${\hat T}_{\ell;a}$ for $a>0$.

Note that we can relate our parameterization to the one proposed in Hu (2001);
If we reexpress equation (\ref{4mom}) as 
\begin{equation}
\langle a_{\ell m_1} a_{\ell m_2} a_{\ell m_3} a_{\ell m_4}\rangle
=\sum_{\alpha=0}^{\ell}{\bar T}_{\ell;\alpha}{\bar {\cal T}}^{\alpha;\ell}_{m_1m_2m_3m_4} 
\end{equation}
where ${T}_{\ell;a}={\cal L}^{a\alpha}_{\ell}{\bar T}_{\ell;\alpha}$ 
then $Q^{\ell \ell}_{\ell \ell}$ as defined in equation 15 of
Hu (2001) can be written as
\begin{eqnarray}
Q^{\ell \ell}_{\ell \ell}(2\alpha)={\bar T}_{\ell;\alpha}+
2(4\alpha+1)\sum_{\beta} 
\left\{ \begin{array}{ccc} \ell & \ell & 2\alpha \\
 \ell & \ell & 2\beta \end{array}  \right\}{\bar T}_{\ell;\beta} .
\end{eqnarray}

The numerical implementation of these estimators is more involved
than for the bispectrum. If we omit the numerous symmetries,
we have to consider for each $\ell$
a set of up to $8\ell^3$ Wigner 3J symbols (compared to
just one for the bispectrum). There are reasonably fast ways
for constructing the Wigner 3J symbols (Schulten and Gordon 1976) but the number
of operations per estimator scales as ${\cal O}(\ell^6)$. 
For repeated computations of the estimators (eg. in Monte Carlo studies),
this can partially be avoided by
storing the precomputed estimators in a lookup table, with
the amount of memory required 
scaling as ${\cal O}(\ell^4)$. 

Clearly, to be able to estimate the trispectrum
on small angular scales, approximate methods must be developed
to make the procedure computationally feasible.
However, the ability to constrain non-Gaussianity on large angular
scales is in any case more important physically for two reasons;
the ratio of the non-Gaussian to the Gaussian signal will
in general be higher for lower moments, and
the signal to noise is better for low $l$.
To understand these points, let us assume a source for non-Gaussianity which
leads to approximately scale invariant moments of the gravitational
potential on arbitrary scales. i.e. $\langle \Phi(R)^N\rangle$ is 
constant for any $R$,
where $\Phi(R)$ is the gravitational potential within a ball of
radius $R$ and $\langle \cdots \rangle$ denotes the ensemble average.
This might be expected from a primordial source with no preferred
scale such as inflation (Komatsu \& Spergel 2000) 
or from an active source where the only 
scale is set by the horizon today (Durrer {\it et al} 2000). 
Current observations of the CMB certainly favor such
scale-invariant descriptions of the potential.
One then expects the moment of order $N$ of the $a_{\ell m}$ to scale
as $\ell^{2(1-N)}$. This signal will be competing against the
fluctuations due to the disconnected (or Gaussian) part, which is
proportional to  $N!\ell^{-(2N+1)/2}$, the former therefore dominating
for $N>2$. Since the power spectrum for white noise has constant amplitude,
the signal to noise as a function of scale will have the same form as
the scale invariant power spectrum itself, therefore being
larger for smaller $l$, ie. larger angular scales.

\section{Results}
\label{res}
As an application of the formalism described in Section
\ref{est}, we estimate the trispectrum of the
coadded 53 and 90 GHz $COBE$-DMR 4 year sky maps in HEALPix format
(G\'{o}rski {\it et al} 1999). The resolution of the maps
is $N_{\mbox{side}}=64$ or $49152$ pixels. 
We do not extend our analysis beyond $\ell_{\mbox{max}}=20$
since the signal to noise is poor for higher $l$.
Hence the maximal number of independent non-Gaussian estimators
for the trispectrum is ${\rm int}(\ell_{\mbox{max}}/3)=6$.
We set the pixels in the extended Galactic
cut (Banday {\it et al} 1997) to zero and subtract the residual monopole and
dipole of the resulting map. After convolving the maps
with spherical harmonics to extract a set of
$a_{lm}$'s for $l \leq 20$ we then apply equation (\ref{tnull}).
To validate our software, 
we have estimated the bispectrum of the $COBE$-DMR 4 year sky data
repixelized in the HEALPix format (for convenience denoted by {\sc ec}) 
and reproduced the results of Ferreira, Magueijo \& G\'{o}rski (1998),
and in particular the strong non-Gaussian
signal present at $\ell=16$. When an equivalent map, from which
that part of the DMR time stream contaminated by the 
`eclipse effect\footnote{The \lq eclipse effect' was an orbitally
modulated signal which took place for approximately two months every year 
around the June solstice when the $COBE$ spacecraft repeatedly flew through the
Earth's shadow.}' 
is removed (denoted {\sc nec}), is subsequently analyzed 
we also reproduce the results of Banday, Zaroubi
\& G\'{o}rski 2000, namely that the non-Gaussian signal is no longer detected.
For our subsequent analysis we will present the trispectra of
{\it both} the {\sc ec} and {\sc nec} data.

One of our primary concerns is to compare our results with 
the assumption that the CMB sky measured by $COBE$-DMR is Gaussian. To do
so, we generate 10000 full-sky
maps at the same resolution using a scale invariant power spectrum normalized to 
$Q_{rms-PS}=18\mu$K (G\'{o}rski {\it et al} 1998).  
We convolve each map with the DMR beam and add uncorrelated
pixel noise with rms amplitude $\sigma_n = 15.95 \mathrm{mK} / \sqrt{N_{obs}}$,
(where $N_{obs}$ is the number of times a given pixel was observed);
we then subject the synthetic map to the same procedure
as the original data.

\vbox{\vskip -0.cm \hskip -0.5cm\epsfxsize=9.cm\epsfbox{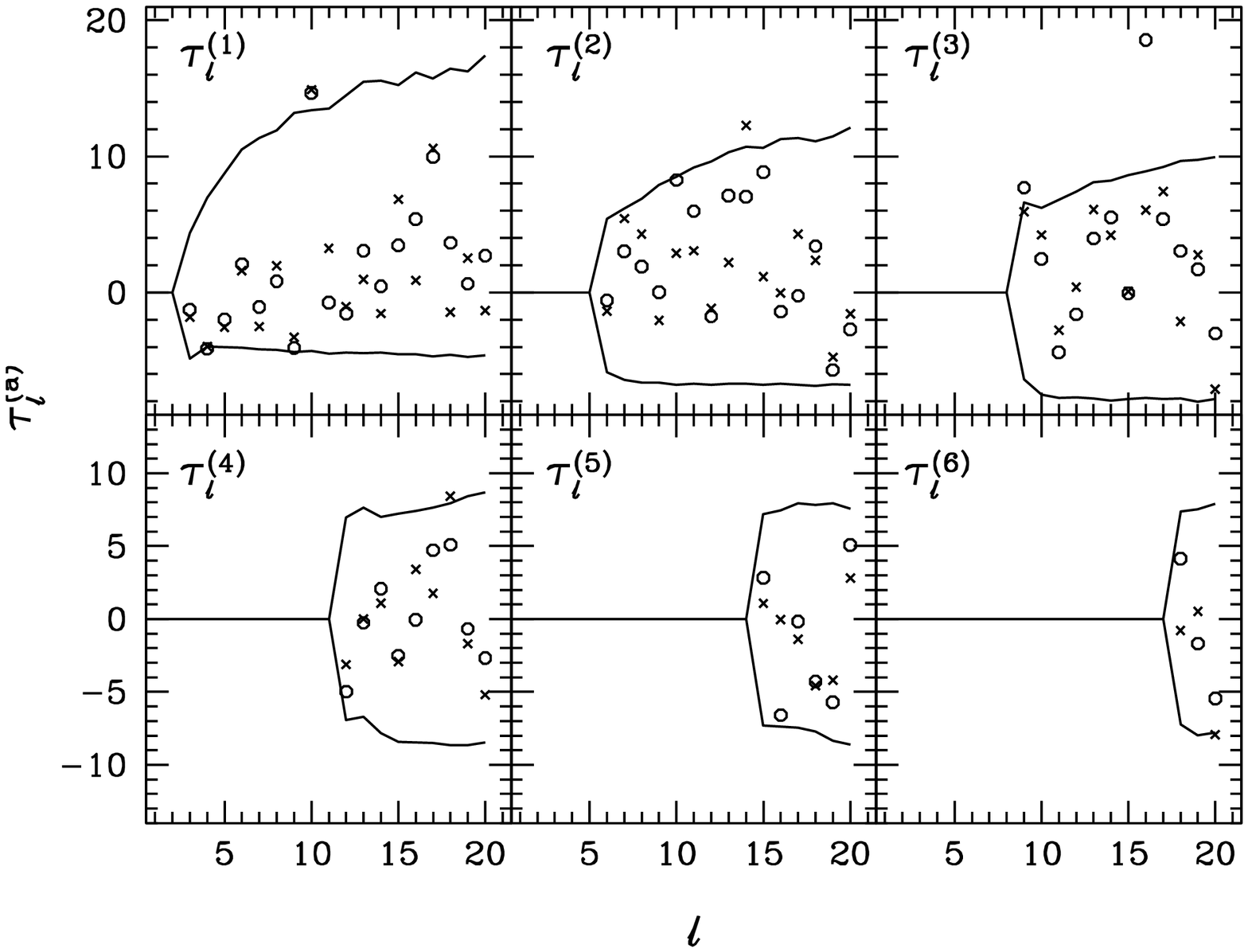}} 
\vskip 0.1cm
{ \small
  F{\scriptsize IG}.~1.--- The six estimators
of the normalized trispectrum applied to the
{\sc ec} data (circles) and the {\sc nec} data (crosses). 
95\% of all simulated Gaussian 
skies lie within the solid lines. Although removing the `eclipse'
data changes the noise properties, we find that the Gaussian confidence
limits essentially remain unchanged.
\label{fig2}}
\vskip 0.25cm

Figure 1 shows the trispectra of the DMR data together 
with Gaussian 95\% confidence limits.
Instead of the ``raw'' estimator (\ref{tnull}) we prefer to use the 
normalized trispectrum,
$\tau^{(a)}_\ell={\hat T}_{\ell;a}/{\hat C}^2_{\ell}$ for $a>1$ (where 
${\hat C}_{\ell}=\frac{1}{2\ell+1}\sum_{m}|a_{\ell m}|^2$),
thus effectively removing the dependence on the power spectrum.
This prevents fluctuations in the power spectrum from introducing
spurious signals and from masking real non-Gaussianities.
Figure 1 shows that in this case, most values fall within the $95\%$ 
confidence lines and demonstrate the scatter expected for a
Gaussian random field.

Of particular interest is the value of the normalized $\tau^{(3)}$ at 
$\ell=16$ in figure 1. One finds that $99.9 \%$ of the Gaussian
models in the {\sc ec} case have a smaller $\tau^{(3)}$ than the
measured one. This is clearly a manifestation of the non-Gaussianity found
in Ferreira, Magueijo \& G\'{o}rski (1998) which is highly localized
in $\ell$ space. However, if we estimate $\tau^{(3)}$ for the {\sc nec}
we find that it falls comfortably within the $95\%$ confidence limits.
This leads us to believe that this detection of non-Gaussianity results
from the `eclipse effect', 
consistent with the hypothesis of Banday, Zaroubi \& G\'{o}rski (1999).

\vbox{\vskip -0.cm \hskip -0.5cm\epsfxsize=9.cm\epsfbox{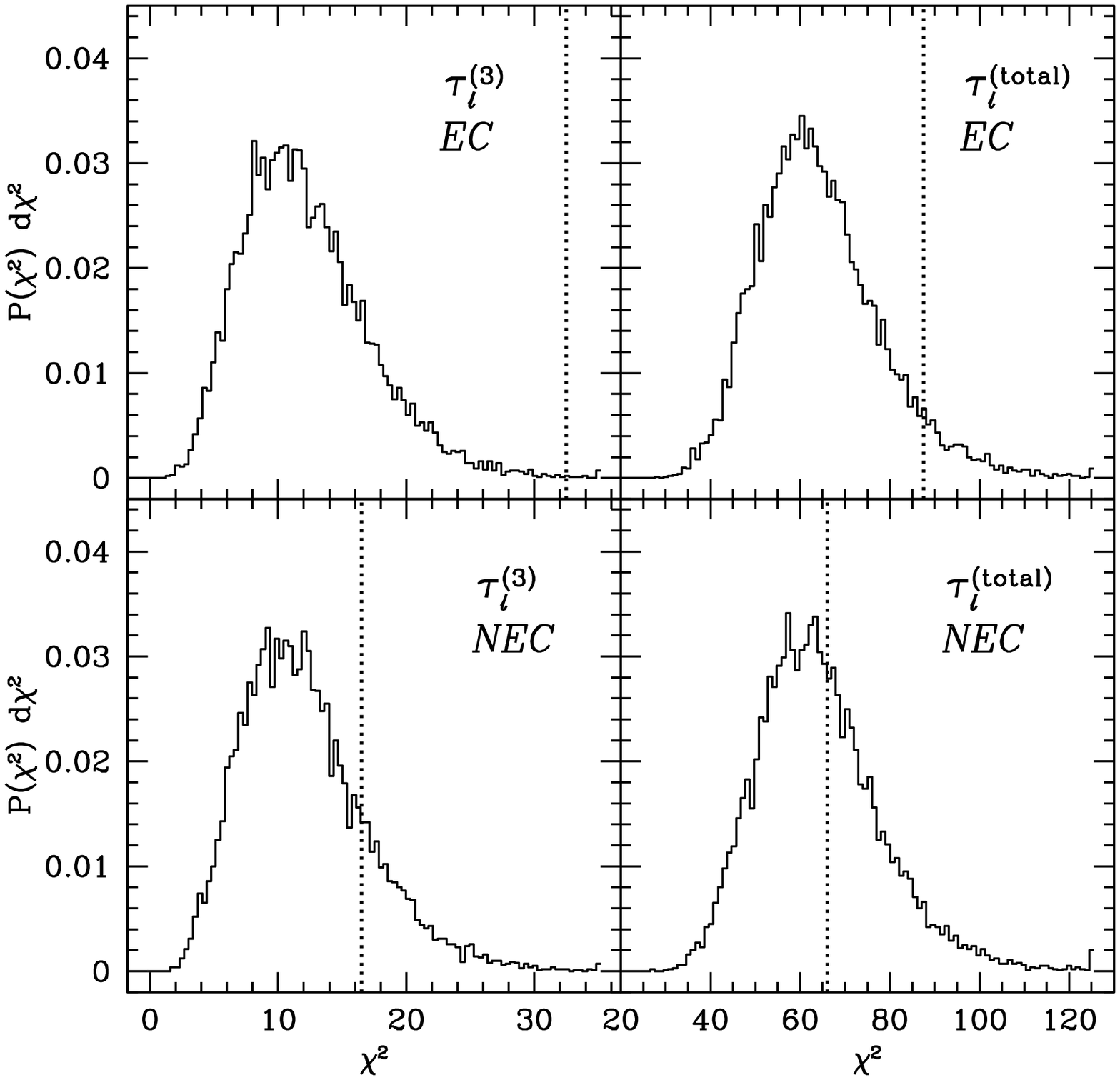}} 
\vskip 0.1cm
{ \small
  F{\scriptsize IG}.~2.--- The $\chi^2$ distribution of the
Gaussian models (histogram) and the actual data value (dotted line)
for the {\sc ec} (top graphs) and {\sc nec} (bottom graphs) datasets.
The left two graphs show $\tau^{(3)}$ which contains the main
contribution to the non-Gaussian signal and the right graphs
show the total $\chi^2$ over all six non-Gaussian estimators,
$\tau^{(1)}$ to $\tau^{(6)}$.
\label{fig3}}
\vskip 0.25cm

Let us now construct a goodness of fit for our statistic. In
Ferreira, Magueijo \& G\'{o}rski (1998), a modified $\chi^2$ was
constructed which took into account the non-Gaussian distribution
of each method: as above, the distribution of each estimator for
a Gaussian sky was constructed and used as an approximate
likelihood function to evaluate the goodness of fit. One
shortcoming of such a method was that correlations between
the estimates for different $\ell$s  were discarded. 
To include them, we use the Gaussian ensemble of data sets to derive the
expectation values $<>_G$ and the covariance matrix $C$  for both
the power spectrum, $C_\ell$, and all seven trispectrum
estimators, $\tau^{(0)}_\ell$ to $\tau^{(6)}_\ell$. We
proceed to calculate the $\chi^2$ value for the estimator
$\EE$ and the data set $\DD$,
\begin{equation}
\chi^2 [\EE,\DD] \equiv \sum_{\ell,\ell'}
\left(\left< \EE_\ell \right>_{\mathrm{G}} - \EE[\DD]_\ell\right)
C^{-1}_{\ell\ell'}
\left(\left< \EE_{\ell'} \right>_{\mathrm{G}} - \EE[\DD]_{\ell'}\right) ,
\end{equation}
using as data sets the {\sc ec} data and the {\sc nec} data.
Finally we use  another 10000 Gaussian realizations to estimate the
expected distribution of the $\chi^2$ for both the {\sc ec} and the
{\sc nec} data.


For all normalized non-Gaussian trispectrum estimators ($\tau^{(1)}$
to $\tau^{(6)}$) we find that 94\% of the Gaussian models
have a smaller $\chi^2$ than the {\sc ec} data as can be 
seen in figure 2.
As expected the main contribution to the $\chi^2$ for the
{\sc ec} data stems from $\tau^{(3)}$ at $\ell = 16$;
indeed, this is the only 
normalized trispectrum estimator which exhibits any significant
non-Gaussianity, in this case at about 99.9\%.
If we use the {\sc nec} data, the detection
vanishes. In this case, 60\% of all Gaussian models have a lower
$\chi^2$ when computed over all six trispectrum estimators
(83\% for $\tau^{(3)}$ alone). Hence the {\sc nec} data
is compatible with Gaussianity.

\section{Discussion}
\label{dis}
In this paper, we have derived an estimator for the trispectrum
of a scalar random field on the sphere. Application of this
estimator, normalized by the power spectrum
(a procedure adopted in Ferreira, Magueijo \& G\'{o}rski, 1998
for the bispectrum, see also Komatsu {\it et al} 2002 for
a detailed discussion), to the $COBE$-DMR data
provides evidence for non-Gaussianity at the 94\% confidence level.
As in the case of the bispectrum, the signal is mainly
present in the $\ell=16$ multipole (and the $\tau^{(3)}$ estimator here).
However, when data is excluded to correct for the `eclipse effect', 
the non-Gaussian behavior is removed, allowing us to conclude that
the non-Gaussianity present in the uncorrected sky maps is not
cosmological in origin.

The detection of a signal that is so strongly localized 
in $\ell$ space provides convincing support to our contention
that the trispectrum is an important and sensitive probe
of non-Gaussianity in the frequency (scale) domain. It affords
complementary information to the bispectrum since it is an even
moment, and, despite the higher computational effort required,
has the obvious advantage in that it can probe all values of
$\ell$, not just the even ones. 

Interestingly enough, from a theoretical perspective, 
there may be some possible sources of non-Gaussianity
for which the trispectrum provides a far more sensitive test 
than the bispectrum. In many
cases a given moment of the $a_{\ell m}$s can be expressed as the
projection of a cosmological field. If that field is vector-like 
in nature (as in the case of the Doppler effect or the Ostriker-Vishniac effect 
and its non-linear extensions), any
odd moment may suffer from the Sunyaev-Kaiser cancellation, where
the integral of a given wavenumber, $k$, over a smoothly varying projection
function with width $\sigma$ tends to suppress the moment by a factor 
of order $1/(\sigma k)^2$ (Sunyaev 1978, Kaiser 1985, Scannapieco 2000). 
For even moments one can always construct a
scalar component which will not be subject to this cancellation.
Such a tool will be of great use in the analysis of the data sets 
from the MAP and Planck Surveyor satellites.

\section{Acknowledgments}

We gratefully acknowledge use of the HEALPix software package
in this publication (see {\tt http://www.eso.org/science/healpix/}).
We thank James Binney, Carlos Contaldi, Michael Joyce, Janna Levin
and Jo\~ao Magueijo
for stimulating discussions. MK acknowledges
financial support from the Swiss National Science Foundation
under contract 83EU-062445. PGC is supported by the Funda\c{c}\~{a}o
Ci\^encia e Tecnologia. PGF thanks the Royal Society for 
support.

{\it Note added in proof} -- E. Komatsu investigates the trispectrum of the
$COBE$ DMR data in his Ph. D. thesis. His conclusions agree with ours, namely
that the $COBE$ data is consistent with Gaussian initial fluctuations.

\end{document}